Enhancement of Critical Current Densities in (Ba,K)Fe$_2$As$_2$ by 320 MeV Au Irradiation in Single Crystals and by High-Pressure Sintering in PIT Wires

Sunseng Pyon[1], Toshihiro Taen[1], Fumiaki Ohtake[1], Yuji Tsuchiya[1], Hiroshi Inoue[1], Hiroki Akiyama[1], Hideki Kajitani[2], Norikiyo Koizumi[2], Satoru Okayasu[3], and Tsuyoshi Tamegai[1]

[1]Department of Applied Physics, The University of Tokyo, Hongo, Bunkyo, Tokyo 113-8656, Japan

[2]Fusion Research and Development Directorate, Japan Atomic Energy Agency (JAEA), 801-1 Mukoyama, Naka, Ibaraki 311-0193, Japan

[3]Advanced Science Research Center, Japan Atomic Energy Agency (JAEA), Ibaraki 319-1195, Japan

We demonstrate a large enhancement of critical current density ($J_c$) up to $1.0 \times 10^7$ A/cm$^2$ at 5 K under self-field in (Ba,K)Fe$_2$As$_2$ single crystals by irradiating 320 MeV Au ions. With the very promising potential of this material in mind, we have fabricated a (Ba,K)Fe$_2$As$_2$ superconducting wire through a powder-in-tube method combined with the hot isostatic pressing technique, whose effectiveness has been proven in industrial Bi2223 tapes. The $J_c$ in the wire at 4.2 K has reached 37 kA/cm$^2$ under self-field and 3.0 kA/cm$^2$ at 90 kOe. Magneto-optical imaging of the wire confirmed the large intergranular $J_c$ in the wire core.

The discovery of superconductivity in the iron-oxipnictide superconductor LaFeAsO$_{1-x}$F$_x$ [1] has elicited great excitement and generated enormous amount of research activities. Many iron-based superconductors (IBSs) exhibit high critical temperatures ($T_c$'s) [2-5], large upper critical fields ($H_{c2}$'s) [6-11], and relatively small anisotropies compared with cuprate superconductors [8,12,13]. They also show very high critical current density ($J_c$) up to $10^6$ A/cm$^2$ in high magnetic fields. This is almost the same level as $J_c$ of NbTi or Nb$_3$Sn which is practically applied for high-field magnets, so they are very promising for high-field applications [14], e.g. high-performance superconducting wires and tapes. IBSs have attracted great interest in the fields of condensed matter physics and superconducting applications, based on such outstanding characteristics.

One of the fundamental questions in IBSs is how much $J_c$ can be enhanced. The $J_c$ of superconductors is one of the extrinsic parameters that can be enhanced by introducing artificial defects. Irradiations of energetic particles are the well-known methods for this purpose. In particular, protons [15] and heavy ions [16] have been proven to be very effective in enhancing $J_c$ in single crystals of cuprate superconductors. These methods may not be practical, but it is well known that enhancements of $J_c$ by similar defect structures have been implemented in a more economical way in an industrial coated conductor of YBa$_2$Cu$_3$O$_7$ [17]. Similar effects of particle irradiations have also been demonstrated in single crystals of IBSs with heavy ions [18-20] and protons [21,22]. Among various IBSs, 122-type materials have been widely studied for the availability of their high-quality single crystals. Among them, (Ba,K)Fe$_2$As$_2$ with $T_c$ up to 38 K is reported to have the largest $J_c$ (2 x $10^6$ A/cm$^2$ at 5 K and 2 kOe) and it is enhanced up to 5 x $10^6$ A/cm$^2$ at 5 K and 2 kOe after irradiating 1.4 GeV Pb ions with a dose-equivalent matching field of $B_\Phi$ = 210 kOe [19]. By optimizing the condition of irradiation, further enhancement of $J_c$ in IBSs is expected, and the achieved value and behavior of $J_c$ can be a target for practical superconducting wires of IBSs.

In order to bring out the potential of remarkably high $J_c$ of IBSs for practical applications, several studies about superconducting wires of IBSs have been performed soon after the discovery. The best candidates for applications in the IBS family are the 122-type compounds, such as (*AE*,K)Fe$_2$As$_2$ (*AE* = alkali earths). They have small anisotropies of 2–3, moderate $T_c$'s, large $H_{c2}$'s, and large $J_c$'s [13]. The $H_{c2}$ for (Ba,K)Fe$_2$As$_2$ exceeds those in both MgB$_2$ and Nb$_3$Sn at low temperatures, so (Ba,K)Fe$_2$As$_2$ attracts great technological interest [11,23]. In order to use 122-type materials for superconducting wires, growth processes of polycrystalline samples and wire fabrication techniques should be improved. Some studies for IBS wires suggest that weak links between superconducting grains are the main reason for the low transport $J_c$ and its strong field dependence. The performance of the wires has been much improved by several methods, the combination of several times of cold press and hot press [24], addition of Ag, Pb, and Sn [25-30], or by texturing tapes [27,29,30], and so on.

In order to develop the superconducting wire of IBSs for practical applications, $J_c$ of the core material should be high and weak links between superconducting grains should be suppressed by optimizing the fabrication process. In this paper, we try to answer two fundamental questions in IBSs; (1) how much can $J_c$ be enhanced ultimately, and (2) what is the most important process in making high-performance superconducting wires of IBSs. Answers to these two questions can be a good target and guiding principle for making practical wires of IBS. For this purpose we show the significant enhancement of $J_c$ both in single crystal and in superconducting wires of (Ba,K)Fe$_2$As$_2$ as a promising materials with high $J_c$. We demonstrate a strong enhancement of $J_c$ up to 1.0x10$^7$ A/cm$^2$ at 5 K under self-field in (Ba,K)Fe$_2$As$_2$ single crystals by 320 MeV Au irradiation. We also demonstrate the enhancement of transport $J_c$ up to 3.7x10$^4$ A/cm$^2$ at 4.2 K under self-field in (Ba,K)Fe$_2$As$_2$ powder-in-tube (PIT) wires by using only the hot isostatic pressing technique (HIP) without other special treatments such as adding other metals or applying the texturing. The HIP process has been proven to be effective in enhancing $J_c$ of industrial Bi2223 tapes [31-33].

Single crystals of (Ba,K)Fe$_2$As$_2$ were synthesized by the self-flux method using FeAs flux. We used Ba pieces, K ingots, and FeAs powder as starting materials. FeAs was prepared by placing stoichiometric amounts of As pieces and Fe powder in an evacuated quartz tube and reacting them at 700 $^o$C for 40 h after heating them at 500 $^o$C for 10 h. A mixture with a ratio of Ba:K:FeAs = 0.6:0.44:4 was placed in an alumina crucible and sealed in a stainless steel container [34] in a nitrogen-filled glove box, and heated for 1 h at 1100 $^o$C followed by cooling to 900 $^o$C at a rate of 5$^o$C/h. 320 MeV Au ions were irradiated into (Ba,K)Fe$_2$As$_2$ along the *c* axis using the tandem accelerator in JAEA to create columnar defects at dose-equivalent matching fields of $B_\Phi$ = 10 and 80 kG. (Ba,K)Fe$_2$As$_2$ superconducting wires were fabricated by the ex situ PIT method. Polycrystalline samples of (Ba,K)Fe$_2$As$_2$ were prepared by the solid-state reaction. A mixture with a ratio of Ba:K:FeAs = 0.6:0.44:2 was sealed in the same way as single-crystal synthesis. It was heated for 5 h at 1,100$^o$C after soaking for 5 h at 600 $^o$C. The prepared (Ba,K)Fe$_2$As$_2$ polycrystalline sample was ground into a fine powder in the nitrogen-filled glove box. The ground powder was filled into a silver tube with an outer diameter of 4.5 mm and an inner diameter of 3 mm, then cold drawn into a square wire with a diagonal dimension of about 1.2 mm. After cutting the drawn wire into ~3 cm pieces, each piece was put into a 1/8 inch copper tube and redrawn into the same size as the silver sheathed wire. Both ends of the wire were sealed using an arc furnace for HIP treatment. They were finally heated for 4 h at 600 °C in Ar atmosphere under a pressure of 120 MPa.

The phase identification of the sample was carried out by means of powder X-ray diffraction (M18XHF, MAC Science) with Cu-K$\alpha$ radiation. Bulk magnetization was measured using a superconducting quantum interference device (SQUID) magnetometer (MPMS-5XL, Quantum Design). Resistivity and current–voltage ($I$–$V$) measurements up to 90 kOe were performed by the four-probe method with silver paste for contacts. $I$–$V$ measurements were performed in a bath-type cryostat (Spectromag, Oxford Instruments). In order to characterize the composition of $(Ba,K)Fe_2As_2$, scanning electron microscopy and energy dispersive X-ray spectroscopy (SEM-EDX) was used. For MO imaging, the wire was cut and the transverse cross sections were polished with a lapping film. An iron-garnet indicator film is placed in direct contact with the sample, and the whole assembly was attached to the cold finger of a He-flow cryostat (Microstat-HR, Oxford Instruments). MO images are acquired by using a cooled CCD camera with 12-bit resolution (ORCA-ER, Hamamatsu).

The $J_c$ characteristics of the pristine and 320 MeV Au irradiated $(Ba,K)Fe_2As_2$ single crystals are summarized in Figs. 1(a) and 1(b). $J_c$'s are evaluated from the irreversible magnetization by using the extended Bean model [18]. Figure 1(a) shows $J_c$ as a function of magnetic field along the $c$-axis at 5 K for the pristine and two irradiated crystals ($B_\Phi$ = 10 kG and 80 kG). $J_c$ of the pristine crystal is about 2.0 x 10$^6$ A/cm$^2$ at 5 K under self-field, which is comparable to that reported in ref. 19. It shows a significant increase up to 1.0 x $10^7$ A/cm$^2$ by 320 MeV Au irradiation with a dose of $B_\Phi$ = 80 kOe. This $J_c$ value is twice larger than that for 1.4 GeV Pb-irradiated crystal ($B_\Phi$ = 210 kOe) [19], and one of the largest values so far reported in IBS single crystals. The reason why low-energy irradiation (320 MeV) of Au-ion is more effective than high-energy irradiation (1.4 GeV) of Pb-ion for increasing $J_c$, is not clear yet. Similar particle energy dependence of $J_c$ has been observed in Au-irradiated Ba(Fe,Co)$_2$As$_2$ [35]. One of the possibilities is that defects are not parallel to each other but become splayed at lower ion energies by the sample stopping power, which was observed in Bi$_2$Sr$_2$CaCu$_2$O$_{8+y}$ [36]. Splayed structure is more effective to suppress the motion of vortices than only parallel structure since magnetic relaxation via half-loop excitation becomes slower by forced entanglement of vortices, resulting in enhancement of $J_c$ [37]. The remarkably high $J_c$ value barely depends on the magnetic field, and it sustains the value of 7.0 x $10^6$ A/cm$^2$ even at 40 kOe. The irradiated (Ba,K)Fe$_2$As$_2$ crystal also retains the high $J_c$ value of 1.0 x $10^6$ A/cm$^2$ even at 30 K below 35 kOe. The extremely high $J_c$ and its retention at high magnetic fields and high temperatures indicate that (Ba,K)Fe$_2$As$_2$ is one of the promising candidates for high-field applications, which may replace NbTi and Nb$_3$Sn in the near future.

With the very promising characteristics of (Ba,K)Fe$_2$As$_2$ single crystals described above, we fabricated and characterized PIT wires of (Ba,K)Fe$_2$As$_2$. First, polycrystalline samples for superconducting wires were characterized before the fabrication of the superconducting wires. Figure 2 shows the X-ray diffraction pattern of an as-prepared (Ba,K)Fe$_2$As$_2$ polycrystalline sample. The diffraction pattern has strong peaks of 122 phase indicating that the reaction to 122 phases is complete. Peaks from impurity phases such as FeAs are not detected in the measured pattern. For the (Ba,K)Fe$_2$As$_2$ polycrystalline sample, the formation of superconducting phase was confirmed by the signal of diamagnetism as shown in the inset of Fig. 2. The shielding volume fraction for the $(Ba,K)Fe_2As_2$ polycrystal reached about 110% at 5 K. The onset $T_c$ of $(Ba,K)Fe_2As_2$ is approximately 38 K. This indicates that the composition of the obtained $(Ba,K)Fe_2As_2$ is almost optimal with a potassium content of $x \sim 0.4$, where $T_c$ is 38 K [4,38].

Figure 3(a) shows the $E$-$J$ characteristics of the HIP-processed PIT wire of $(Ba,K)Fe_2As_2$ under different magnetic fields at 4.2 K. Transport $J_c$ as a function of magnetic field at 4.2 K for the PIT $(Ba,K)Fe_2As_2$ wire is shown in Fig. 3(b). Here, we adopt $E = 1$ μV/cm as a criterion to define transport $J_c$ for the $E$-$J$ curves. The transport $J_c$ at 4.2 K has reached 37 kA/cm$^2$ under self-field and 3.0 kA/cm$^2$ at 90 kOe in the $(Ba,K)Fe_2As_2$ PIT wire. $J_c$'s in 122 superconducting wires or tapes from recent publications are also plotted in Fig. 3(b) [24, 28-30]. The self-field $J_c$ in the HIP-processed $(Ba,K)Fe_2As_2$ wire is several times larger than that of $(Ba,K)Fe_2As_2$ PIT wires, which are sintered at ambient pressure [28]. Furthermore, it is almost 10 times larger at high magnetic fields. The obtained value of $J_c$ is roughly the same level as the recently reported $J_c$'s in Sn-added $(Ba,K)Fe_2As_2$ textured tapes [30], is 20-30% of $J_c$ in Sn-added $(Sr,K)Fe_2As_2$ textured tape [29], and is 20-30% of $J_c$ in the HIP-processed wire employing mechanical alloying with a low-temperature process [24]. These results strongly indicate that our HIP treatment for the PIT wire largely contributed to the improvement of $J_c$, although fabrication processes need to be improved.

The HIP-processed wire was further characterized by several physical measurements. First, we measured its magnetization to detect its $T_c$. The temperature dependence of the magnetic moment for the short segment of the wire was measured at an applied field of 10 Oe along the wire direction, and the result is shown in Fig. 4. The onset of the main body of diamagnetism of the (Ba,K)Fe$_2$As$_2$ wire is slightly reduced to 30 K, but the shielding volume fraction is roughly 100%. Subsequently, we performed the SEM-EDX analysis for the wire. The analyzed potassium content $x$, detected from the whole area of the cross section of wires of about 200 x 200 μm$^2$, was 0.43 for Ba$_{1-x}$K$_x$Fe$_2$As$_2$. These values are almost the same as the nominal composition of optimal potassium content $x \sim 0.4$. However, from local area analysis around 10 μm$^2$, the grains that have rich or poor potassium contents were also detected. Furthermore, small amounts of impurities such as FeAs and BaAs were also detected. The distribution of potassium content in 122 phases and impurities may be the main reason for the reduction of $T_c$. To suppress the content of impurities and to obtain optimal 122 phases in wires, fabrication and synthesis processes should be improved.

In IBSs, vortices at grain boundaries are generally pinned more weakly than vortices in the grains, thus the grain boundaries become barriers for current flow. To investigate the quality of the grain boundaries, we performed MO measurements on the PIT wire. Figure 5(a) is an optical image of the transverse cross section in the HIP-processed wire. It is obvious that there is no reaction between the (Ba,K)Fe$_2$As$_2$ core and Ag sheath. The area of the core part was approximately 0.0005 cm$^2$. The thickness of this sample along the wire axis in the optical image is roughly 300 μm. A higher magnification optical image of the core is shown in Fig. 5(b). Small grains of 122 phases with sizes less than 20 μm can be identified. No void is observed, which proves the high density of the wire core caused by high-pressure sintering. Figures 5(c)-5(e) depict MO images of the transverse cross section of the wire core in the remanent state after applying an 800 Oe field along the wire axis for 0.25 s which was subsequently reduced to zero at 4.2, 15, and 25 K, respectively. The bright regions correspond to the trapped flux in the sample. Figures 5(c), 5(d) show the uniform and fully trapped magnetic flux distributions at 4.2 and 15 K in the (Ba,K)Fe$_2$As$_2$ wire core. It guarantees a very uniform bulk current flow in the wire core across many grains. It is clear that weak links across grain boundaries are much improved compared with our previous PIT wire processed at ambient pressure [28]. HIP treatment should contribute to generate strong links between superconducting grains and to enhance intergranular $J_c$. On the other hand, at 25 K, the trapped magnetic flux is inhomogeneous, and several bright spots are visible as shown in Fig. 5(e). These images are slightly similar to the MO images of 1111 and 11 polycrystals and tapes [39,40]. This implies that the intergranular critical current density is smaller than the intragranular critical current density at around $T_c$. This is consistent with the result of magnetization measurement shown in Fig. 3. From the magnetic induction profile, we calculated the intergranular critical current density $J_c^{inter}$. In this calculation, we roughly estimate it as $J_c^{inter} \sim dB/dx$ [41]. Figure 5(f) shows the magnetic induction profiles along the dashed line in Fig. 5(c). $J_c^{inter}$ decreases gradually as the temperature is increased toward $T_c$. $J_c^{inter}$ of 45 kA/cm$^2$ at 4.2 K in the (Ba,K)Fe$_2$As$_2$ PIT wire was obtained from the profiles. These data compare well with the transport $J_c$ shown in Fig. 3(b). All these results allow us to conclude that high-pressure sintering by the HIP process is highly effective for suppressing the weak-link nature of grain boundaries and enhancing the intergranular critical current density in (Ba,K)Fe$_2$As$_2$ wires.

In summary, we have irradiated 320 MeV Au ions into high-quality single crystals of (Ba,K)Fe$_2$As$_2$ and confirmed a significant increase of critical current density $J_c$ up to 1.0x10$^7$ A/cm$^2$ at 5 K under self-field. The very weak magnetic field dependence of $J_c$ in (Ba,K)Fe$_2$As$_2$ suggests that it is one of the best materials for high-field applications. We have also fabricated a (Ba,K)Fe$_2$As$_2$ superconducting wire by the PIT method combined with the hot isostatic press (HIP) technique, and characterized it by X-ray diffraction, magnetization, transport $J_c$, SEM-EDX, and magneto-optical (MO) measurements. Although the quality of the polycrystalline material has not been optimized, MO images show a smooth magnetic induction profile across the wire core, indicating the good improvement of weak links between superconducting grains. The transport $J_c$ at 4.2 K has reached 37 kA/cm$^2$ under self-field and 3.0 kA/cm$^2$ at 90 kOe. It is demonstrated that high-pressure sintering by the HIP technique is highly effective for the enhancement of intergranular $J_c$. Further enhancement of $J_c$ in (Ba,K)Fe$_2$As$_2$ wires is expected by the combination of the HIP technique and other methods, such as the optimization of polycrystal synthesis and texturing wires.

**Acknowledgements**

This work was partially supported by a Grant-in-Aid for Young Scientists (B) (No. 24740238) and the Japan-China Bilateral Joint Research Project by the Japan Society for the Promotion of Science (JSPS).

Fig. 1 (a) Magnetic field dependence of $J_c$ at 5 K for the pristine and 320 MeV Au-irradiated (Ba,K)Fe$_2$As$_2$ crystals. $J_c$ increases up to 1.0 x 10$^7$ A/cm$^2$ in the irradiated crystal at 5 K under self-field. In addition, the magnetic field dependence becomes very weak after irradiation. (b) Temperature dependence of $J_c$ for pristine and irradiated crystals at $H$ = 2 and 35 kOe.

Fig. 2 Powder X-ray diffraction pattern of (Ba,K)Fe$_2$As$_2$ powder. Inset shows zero-field-cooled (ZFC) and field-cooled (FC) magnetizations of (Ba,K)Fe$_2$As$_2$ powder at $H$ = 5 Oe.

Fig. 3 (a) $E$-$J$ characteristic of (Ba,K)Fe$_2$As$_2$ PIT wire under different magnetic fields at 4.2 K. (b) Magnetic field dependences of transport $J_c$'s of the wires. $J_c$'s from recent publications are also plotted [24, 28-30].

Fig. 4 Temperature dependence of magnetization of (Ba,K)Fe$_2$As$_2$ PIT superconducting wire under magnetic field of 10 Oe.

Fig. 5 Optical micrographs of (a) transverse cross section and (b) core region of (Ba,K)Fe$_2$As$_2$ PIT wire. Differential MO images of core region of (Ba,K)Fe$_2$As$_2$ wire in the remanent state at (c) 4.2, (d) 15, and (e) 25 K, after cycling the field up to 800 Oe for 0.25 s. The gray scales spans from -50 to 400 G in (c) and (d), and from -10 to 100 G in (e). (f) Local magnetic induction profiles at different temperatures taken along the dashed lines in (c).

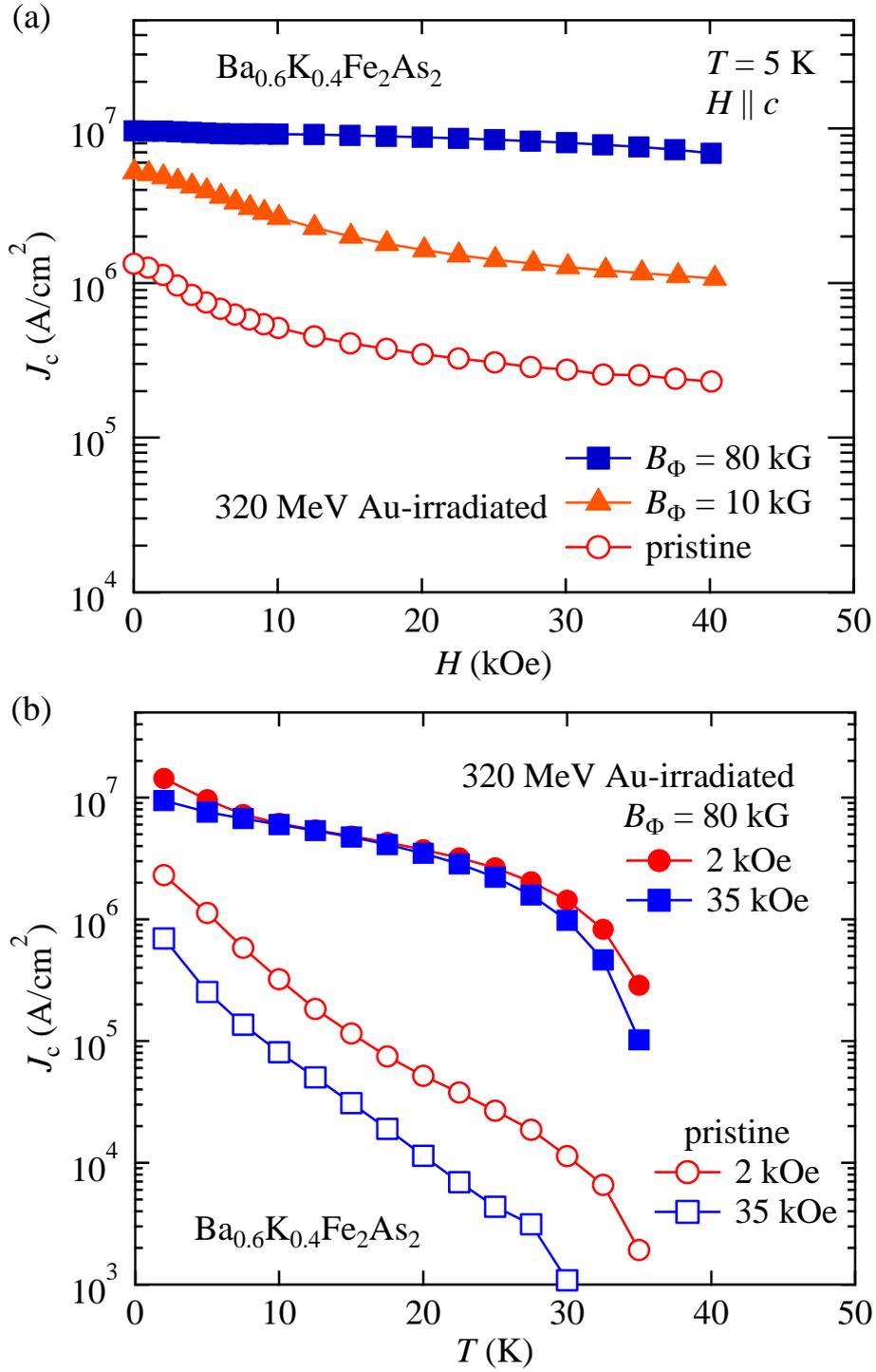

Fig. 1

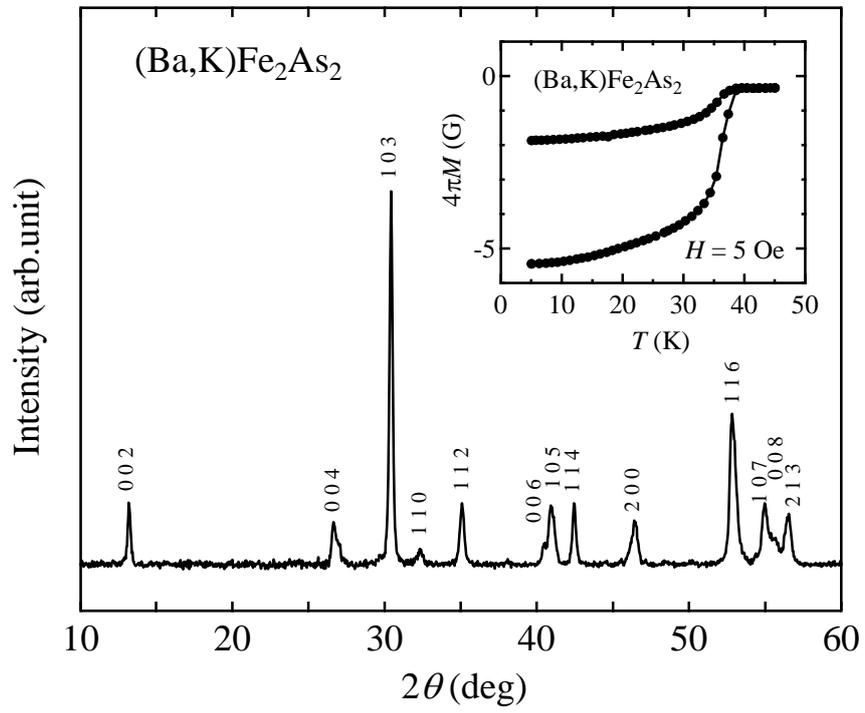

Fig. 2

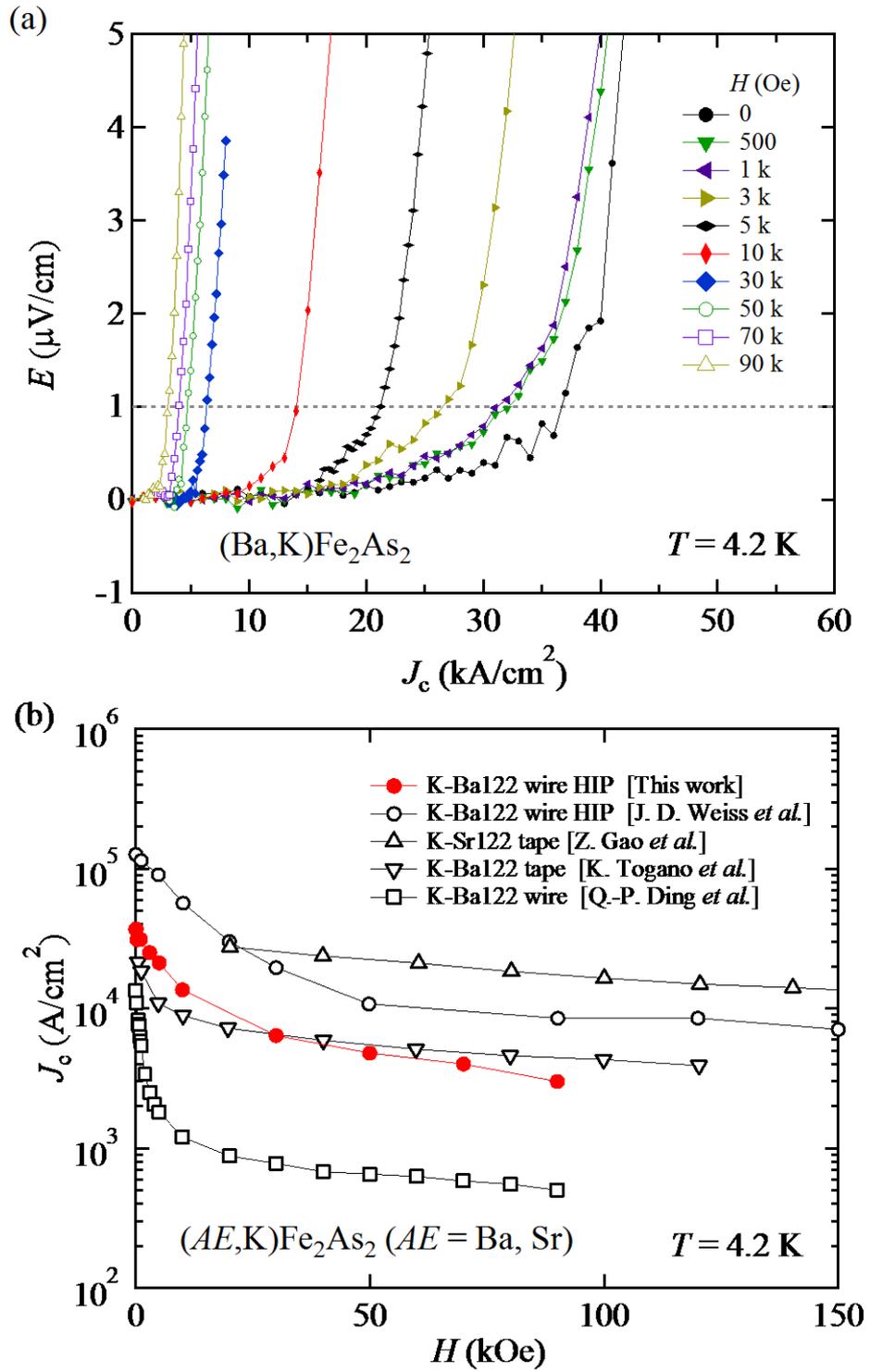

Fig. 3

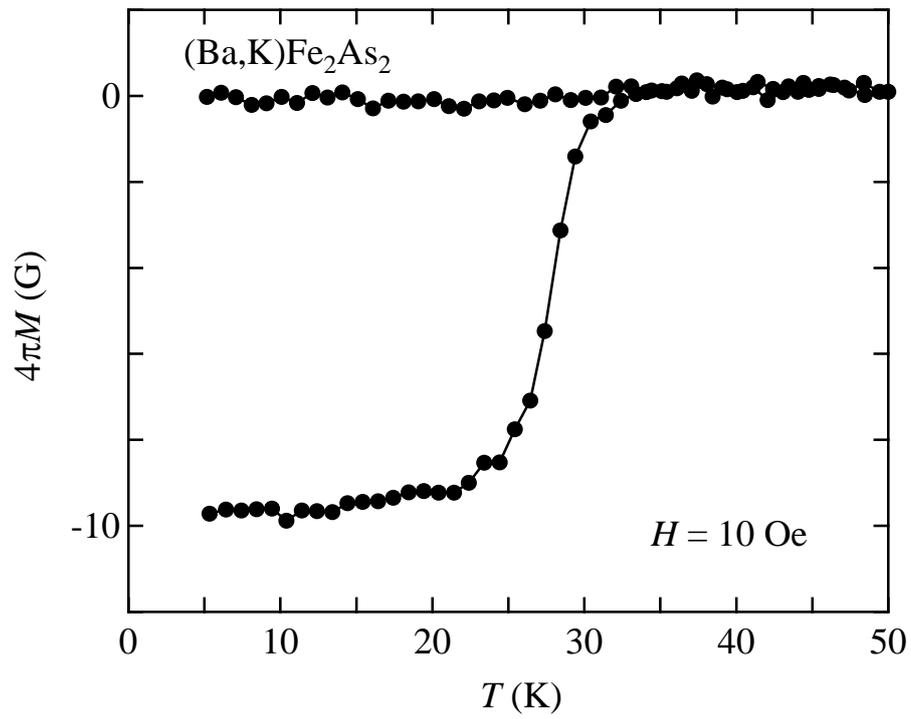

Fig. 4

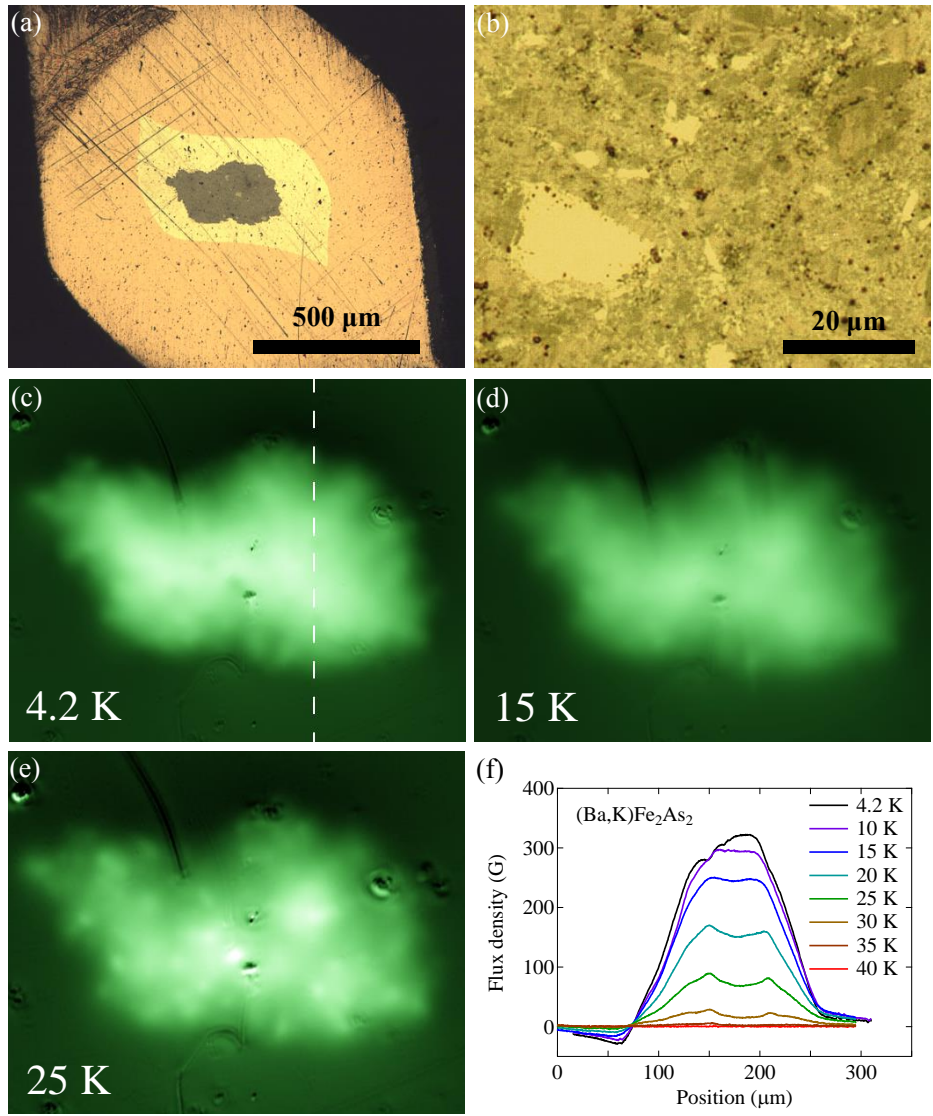

Fig. 5